\documentclass[aps,twocolumn,preprintnumbers,amsmath,amssymb]{revtex4}

\usepackage{graphicx}
\usepackage{dcolumn}
\usepackage{bm}
\usepackage{multirow}
\usepackage{color}
\usepackage{ulem}
\usepackage{amsmath}

\begin{document}
\title{Discrete quantum dot like emitters in monolayer MoSe$_2$:\\
Spatial mapping, Magneto-optics and Charge tuning}

\author{Artur Branny$^{1}$}
\thanks{These authors contributed equally to this work}
\author{Gang Wang$^{2}$}
\thanks{These authors contributed equally to this work}
\author{Santosh Kumar$^{1}$}
\author{Cedric Robert$^{2}$}
\author{Benjamin Lassagne$^{2}$}
\author{Xavier Marie$^{2}$}
\author{Brian D. Gerardot$^{1}$}
\email{b.d.gerardot@hw.ac.uk}
\author{Bernhard Urbaszek$^{2}$}
\email{urbaszek@insa-toulouse.fr}

\affiliation{%
$^1$ Institute of Photonics and Quantum Sciences, SUPA, Heriot-Watt University, Edinburgh EH14 4AS, United Kingdom}
\affiliation{%
$^2$ Universit\'e de Toulouse, INSA-CNRS-UPS, LPCNO, 135 Av. Rangueil, 31077 Toulouse, France}


\begin{abstract}
Transition metal dichalcogenide monolayers such as MoSe$_2$,MoS$_2$ and WSe$_2$ are direct bandgap semiconductors with original optoelectronic and spin-valley properties. Here we report spectrally sharp, spatially localized emission in monolayer MoSe$_2$. We find this quantum dot like emission in samples exfoliated onto gold substrates and also suspended flakes. Spatial mapping shows a correlation between the location of emitters and the existence of wrinkles (strained regions) in the flake. We tune the emission properties in magnetic and electric fields applied perpendicular to the monolayer plane. We extract an exciton g-factor of the discrete emitters close to -4, as for 2D excitons in this material. In a charge tunable sample we record discrete jumps on the meV scale as charges are added to the emitter when changing the applied voltage. 
\end{abstract}


\maketitle
\begin{figure*}
\includegraphics[width=0.82\textwidth]{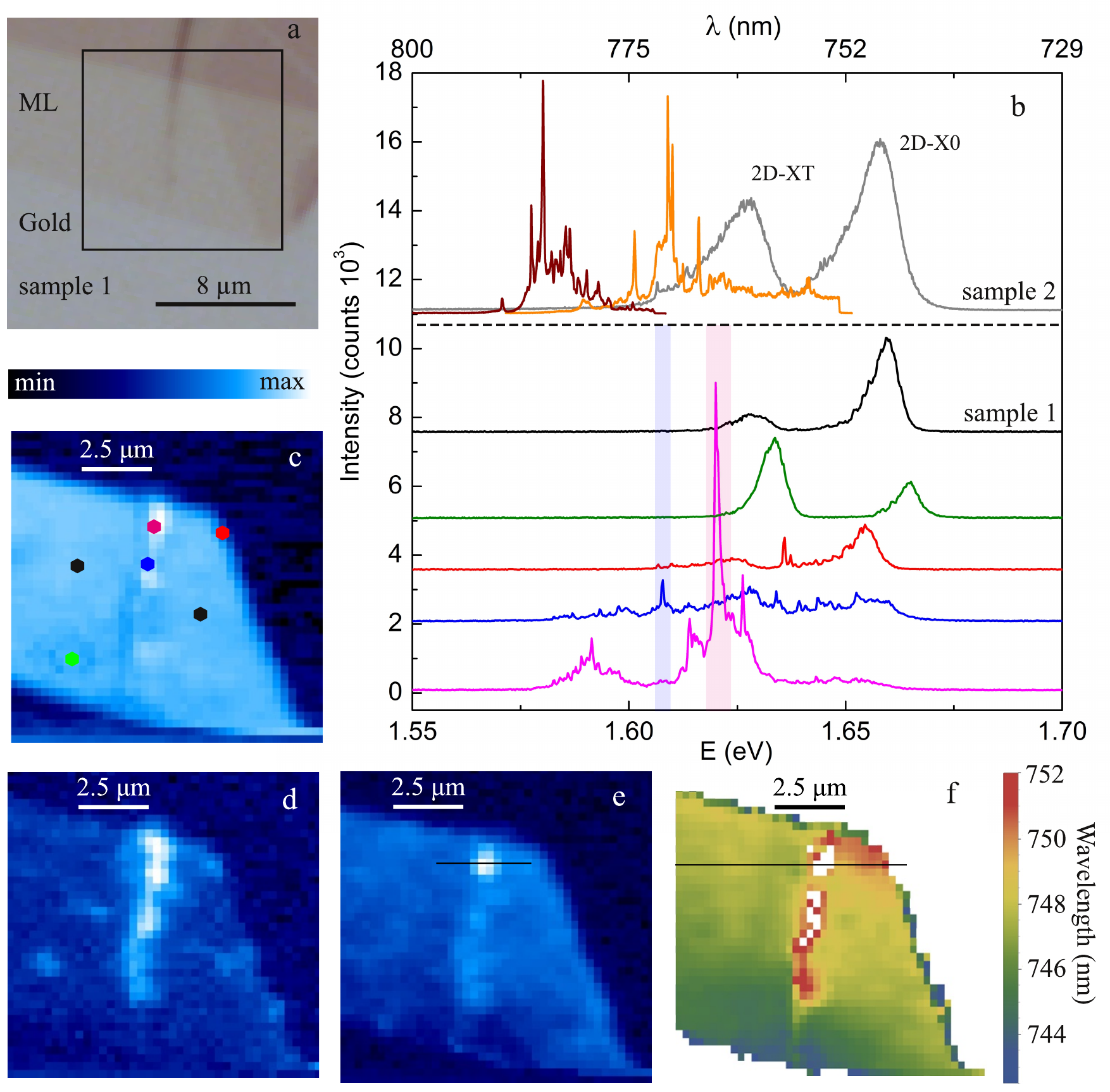}
\caption{\label{fig:fig1} (a) Micrograph (x50 magnification) of sample 1, ML MoSe$_2$ on a gold surface where the flake formed a wrinkle which can be recognised by a line with darker colour. (b) Top 3 spectra from sample 2, all others from sample 1. In black, a typical spectrum showing two- dimensional neutral exciton, 2D-X0 and trion, 2D-XT. In colour, examples of sharp, quantum dot-like emissions found at the edges and along the wrinkle. Space maps of integrated intensity in (c) for the entire spectral window between 683 - 855 nm. Space map of the peak intensity in (d) between 770.56 - 771.91 nm capturing sharp emission from blue (the second from the bottom) spectrum and in (e) between 764.52 - 765.86 nm for emitter from the pink spectrum (the bottom). These locations appear near the wrinkle from the micrograph. (f) Space map of fitted 2D-X0 wavelengths which shows a homogeneous emission apart from the rapid redshift of ~10 meV caused by the wrinkle. The dotted line in  f) includes the pink spectrum at the bottom from b).We show discrete lines at different sample positions, often related to local strain.
}
\end{figure*}
\begin{figure*}
\includegraphics[width=0.9\textwidth]{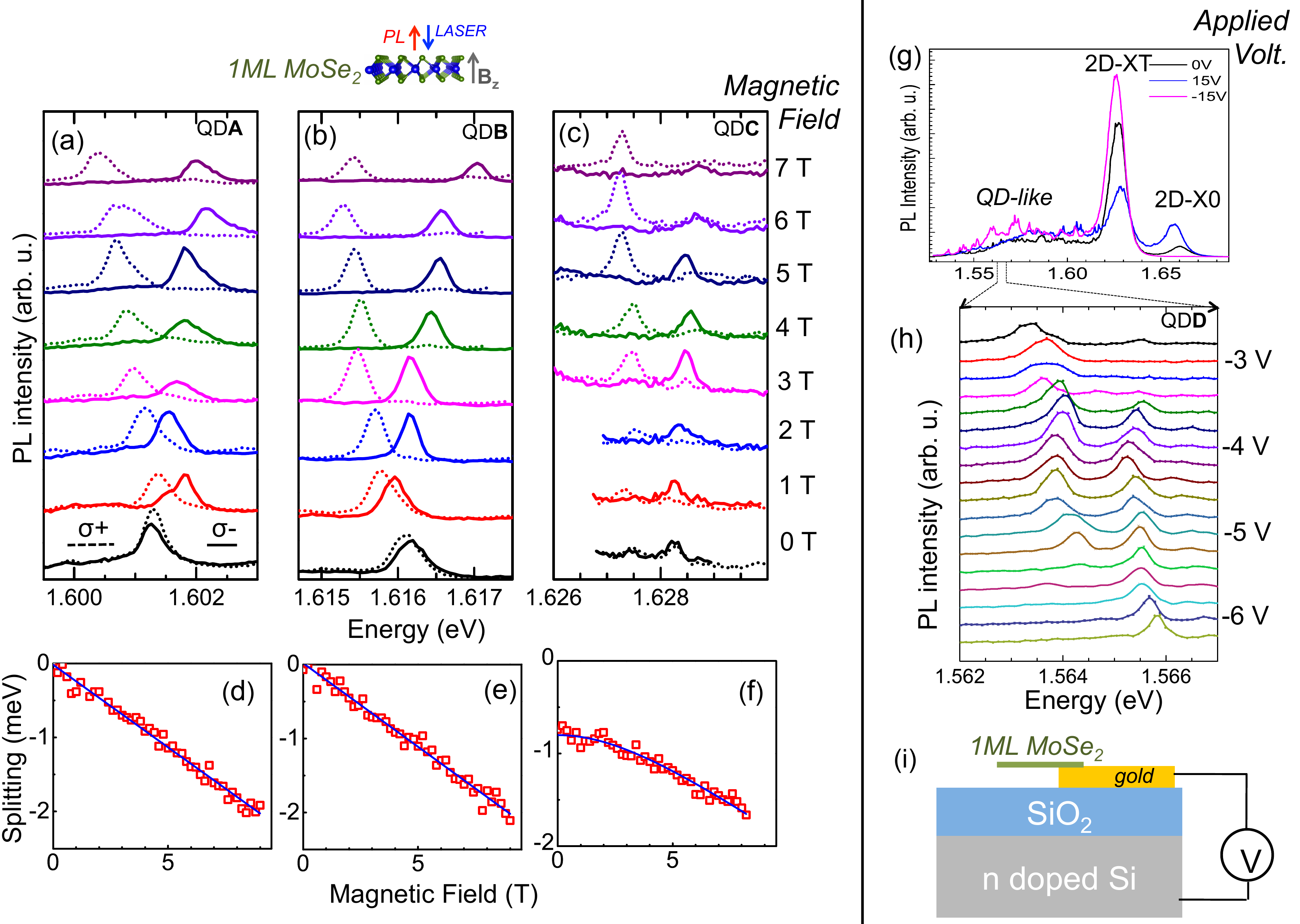}
\caption{\label{fig:fig2}  T=4~K. (a) Photoluminescence spectra from a single discrete line QD A from \textbf{sample 2} taken at different applied magnetic fields $B_z$ perpendicular to the sample plane. Spectra are offset for clarity, solid (dashed) lines are recorded in $\sigma^- (\sigma^+)$ circular detection basis. (b) same as (a), but for emitter QD B. (c) same as (a), but for emitter QD C. (d) Zeeman splitting of QD A emission (red squares), linear fit (blue line). (e) same as (d) but for QD B. (f) As QD C shows a zero field splitting $\delta_1$, here we fit the magnetic field dependence  $\propto\sqrt{\delta_1^2+\Delta_Z^2}$. (g) \textbf{Sample 3}. Tuning the 2D-trion to neutral exciton ratio with the applied bias in sample 3. (h) PL spectra from discrete emitter QD D from charge tunable sample 3 (suspended region) as a function of applied bias voltage. (i) schematic of the charge tunable device.
}
\end{figure*}

\textit{Introduction.---}
Transition metal dichalcogenide (TMD) monolayers (MLs) such as MoS$_2$, MoSe$_2$ and WSe$_2$ have a direct bandgap in the visible region of the spectrum \cite{Mak:2010a,Splendiani:2010a,Ross:2013a}, ideal for optoelectronics applications.
Quantum confinement in three dimensions leading to discrete electronic states in TMD host materials would provide a versatile platform for optical and electrical manipulation of spin and valley states of individual carriers \cite{Kormanyos:2014a,Liu:2014b,Wu:2016a}. The 2D host materials have the advantage of being cost efficient, with highly tunable properties \cite{Ross:2013a,He:2013a} and optical access to the electron valley index in momentum space \cite{Mak:2014a,Xu:2014a}, an additional degree of freedom compared to other solid state qu-bits in III-V quantum dots (QDs) or NV centres in diamond, for example. There are several approaches to achieve 3D quantum confinement, such as patterning TMD MLs \cite{Wei:2015a}, chemically synthesized TMD nano-crystals  \cite{Wilcoxon:1997a,Huang:2000a,Huang:2000b,Chikan:2002a,Etzkorn:2005a,Lin:2013a} and defect engineering \cite{Tran:2016a,Feng:2014a,Tongay:2013a,Zhou:2013a}. \\
\indent In photoluminescence (PL) experiments at T=4~K we observe QD-like, discrete emission lines (FWHM typ.~150-400~$\mu$eV) in energy below the 2D charged exciton (trion) and neutral exciton emission (FWHM typ.~10~meV) in ML MoSe$_2$.  We show that the discrete emission lines stem from spatially isolated regions, linked to the positions of wrinkles in the ML flake. To provide insight into the physical origin of the emitters and to tune their emission properties we perform experiments in magnetic fields applied perpendicular to the ML. We are able to extract an exciton Land\'{e} g-factor of $g= -4$, close to values reported for the 2D neutral excitons in ML MoSe$_2$ \cite{Li:2014a,Macneill:2015a,Wang:2015d}. We also demonstrate discrete charge tuning of the QD-like emitters in a suspended flake. \\
\indent The existence of discrete emitters with intriguing optical properties has been reported very recently in the TMD ML host material WSe$_2$ \cite{Koperski:2015a,Srivastava:2015a,He:2015a,Chakraborty:2015a,Tonndorf:2015a,Kumar:2015a}, but not in MoSe$_2$ to our knowledge. The physical origin of the discrete emitters still needs to be clarified. Between ML MoSe$_2$ used here and WSe$_2$ studied in the literature exist important differences: (i) the QD like emitters can appear background free below the trion PL emission energy, see Figs.~\ref{fig:fig1} and \ref{fig:fig2}, whereas in ML WSe$_2$ the discrete emitters can overlap with broad peaks linked to localized 2D excitons and/or their phonon replica \cite{Dery:2015a}. (ii) The 2D neutral exciton states in ML MoSe$_2$ are optically bright, whereas the lowest energy transition in ML WSe$_2$ is optically dark \cite{Zhang:2015c,Wang:2015e,Dery:2015a,Arora:2015b,Withers:2015a,Echeverry:2016a}, which will impact carrier relaxation and recombination dynamics.
\\
\indent \textit{Samples and Experimental Set-up.---}
The MoSe$_2$ ML flakes are obtained by micro-mechanical cleavage of a bulk crystal using viscoelastic stamping \cite{Gomez:2014a}. Sample 1 and 2 are exfoliated onto a gold substrate. An optical micrograph, in Fig.~\ref{fig:fig1}a, shows sample 1 and marks the region which has been scanned for PL measurements. Despite the highly reflective gold surface, the contrast for MLs and few layered flakes under white light illumination is still noticeable. The dark line penetrating the image from the top is due to a wrinkle which was intentionally created during the transfer process by applying additional shear force. Sample 3 is a charge tunable structure, the MoSe$_2$ ML is suspended between two Cr/Au (5/50 nm) electrodes pre-patterned by E-beam lithography on SiO$_2$(90 nm)/Si substrate. The distance between the two electrodes is 4 $\mu$m.  The  application of a bias voltage between the top Cr/Au electrodes and the n-doped Si substrate (used as a back gate) enables electrical tuning of the resident carrier density. Experiments at $T=4$~K are carried out in a confocal microscope \cite{Kumar:2015a,Wang:2014b}, magnetic fields up to $|B_z|=9~$T can be applied perpendicular to the sample plane i.e. parallel to the light propagation axis (Faraday geometry) \cite{Wang:2015d}. The detection spot diameter is $\leq1\mu$m, i.e. considerably smaller than the ML size of typically $\sim 10~\mu$m$\times10~\mu$m. The PL emission is dispersed in a spectrometer and detected with an Si-CCD camera. The sample is excited with 633~nm or 532~nm lasers.\\
\indent \textit{Results and Discussion.---}
Sample 1 and 2 exfoliated onto gold show for most parts two dimensional neutral (2D-X0) and charged (2D-XT) exciton peaks at usual \cite{Ross:2013a,Zhang:2014a,Wang:2015c,Wang:2015a} energies 1.659 eV and 1.628 eV shown in Fig.~\ref{fig:fig1}b (grey curve for sample 2, black for sample 1).
The FWHM of the 2D exciton PL emission is of the order of 10~meV. In Fig.~\ref{fig:fig1}c, we display a space map of integrated PL intensity for the entire spectral window 730 to 800 nm. This map reveals that the black spectrum from Fig.~\ref{fig:fig1}b is representative for the majority of PL signal coming from the sample with only small variations in intensity, indicating the good overall optical quality of the flake. Remarkably, the 2D exciton emission is not spectrally shifted (within a few meV) compared to samples exfoliated onto SiO$_2$/Si and the trion binding energy of $E_{PL}^\text{2D-X0}-E_{PL}^\text{2D-XT}\approx 31~$meV is not influenced by the choice of metal substrate. Also for the related ML material MoS$_2$ it has been reported that the emission of the direct gap 2D excitons is little affected by the gold substrate \cite{Mertens:2014a,Bhanu:2014a}, presumably due to the very small exciton Bohr radius in ML TMDs \cite{Qiu:2013a,Chernikov:2014a}. The exact ratio of the PL intensities 2D-XT~:~2D-X0 varies across the sample, compare in  Fig.~\ref{fig:fig1}b the 2D exciton spectrum in green and black. This is also observed for the more standard ML MoSe$_2$ on SiO$_2$ system, therefore gate control of the residual carrier density is desirable, as we demonstrate below.\\
\indent At certain locations on the flake (see points marked in Fig.~\ref{fig:fig1}c) we detect spectrally sharp emission lines with typ. FWHM of~$150-400~\mu$eV, similar to those found in WSe$_2$ MLs \cite{Koperski:2015a,Srivastava:2015a,He:2015a,Chakraborty:2015a,Tonndorf:2015a,Kumar:2015a}. The top and the bottom panels of Fig.~\ref{fig:fig1}b show two sets of quantum dot-like PL spectra which were taken from two different ML flakes, sample 1 and 2. The discrete PL signal from MoSe$_2$ is not obscured by broad band emission related to 2D excitons. However, the lines from the QD like emitter in MoSe$_2$ in our samples are closely spaced in energy, making it challenging to filter a single emission line. In time traces we measure spectral jitters up to $\approx$1~meV in some cases, suggesting that the measured linewidth of several hundred $\mu eV$ is also broadened by these fluctuations. We tentatively assign this behaviour to charge noise in the environment of the emitter which could be influenced by the presence of the gold substrate. Also here the integration of QD like emitters in a charge tunable device or simple encapsulation by boron nitride are expected to reduce the jitter in future measurements.\\
\indent Our measurements show that localization of these emitters is correlated with local strain gradients. We generate space maps of integrated PL peak intensities over two narrow spectral windows. Figures~\ref{fig:fig1}d and \ref{fig:fig1}e are space maps for emitters from blue (the second from the bottom) and pink (the bottom) spectra of Fig.~\ref{fig:fig1}b. These figures show that the spatial locations of these emitters coincide with the wrinkle position on the flake from the micrograph (see Fig.~\ref{fig:fig1}a). Localisation lengths of these centres are beyond the diffraction limit of our optical setup. We measure full width at half-maximum (FWHM) of 570 nm for the intense source from Fig. 1e (the bottom spectrum). To understand further the influence of wrinkle we fit peak wavelengths of 2D-X0 emission across the flake. We observe a sudden and sharp redshift as we go across the wrinkle. This shift is equivalent to $\approx$10~meV decrease in energy relative to the average 2D-X0 energy along the solid line. Therefore, as it was shown for WSe$_2$ \cite{Tonndorf:2015a,Kumar:2015a}, QD-like emitters in MoSe$_2$ are localized by sharp and local strain gradients.\\
\indent \textit{Magneto-Photoluminescence.---}
We perform magneto-optics to compare the response of the discrete emitters to the 2D exciton results reported recently for MoSe$_2$ MLs \cite{Li:2014a,Macneill:2015a,Wang:2015d}. In Fig.~\ref{fig:fig2}a to c we show the PL spectrum of typical, discrete lines as a function of the applied magnetic field $B_z$ perpendicular to the ML i.e. parallel to the light propagation axis (Faraday geometry). The discrete emitters QD A and QD B from sample 2 do not show any measurable fine structure splitting at $B_z=0$.
The applied field clearly results in an energy splitting between the $\sigma^+$ and $\sigma^-$ polarized PL components of about $-2~$meV at $B_z=9~$T. We have verified that the energy splitting of the QD like emitters presented in Fig.~\ref{fig:fig2} has the same sign as the 2D neutral exciton in the same MoSe$_2$ ML sample. For QD A and QD B the Zeeman splitting  $\Delta_Z=E(\sigma^+)-E(\sigma^-)$ increase linearly in amplitude with the applied field $B_z$. Fitting the linear function $\Delta_Z=g\mu_B B_z$ to the splitting measured for QD A and QD B we extract a g-factor of  $g=-3.8\pm0.2$ for QD A and  $g=-3.9\pm0.2$ for QD B (where $\mu_B$ is the Bohr magneton). This is very close in value to the g-factor obtained for the 2D neutral exciton and trion in ML MoSe$_2$ \cite{Li:2014a,Macneill:2015a,Wang:2015d}, which hints at a physical relation between the electronic states of the discrete emitters and the 2D-X0 formed by carriers at the direct gap at the K-point.  This is in stark contrast to the reports for localized states in WSe$_2$, that showed g-factors considerably bigger than the values for 2D excitons \cite{Srivastava:2015a,Koperski:2015a,He:2015a,Chakraborty:2015a} with amplitudes between 6 and 13. To understand the origin of these large and variable g-factors and also the value of the 2D-exciton g-factors in ML TMD in gerenral motivates numerous recent studies \cite{Srivastava:2015b,Aivazian:2015a,Wang:2015d,Stier:2016a} and demands future work. Another important difference between the QD like emitters in ML MoSe$_2$ and WSe$_2$ is the absence of the zero field fine structure splitting due to anisotropic electron-hole Coulomb exchange for QD A and QD B in Fig.~\ref{fig:fig2}a and b. For discrete emitters in ML WSe$_2$ investigated in Refs.~\cite{Srivastava:2015a,He:2015a,Chakraborty:2015a} zero field splittings of typically 600~$\mu eV$ have been reported, similar to GaAs and CdSe QDs with reduced symmetry \cite{Gammon:1996a,Kulakovskii:1999a}. So scanning sample 2 carefully for a dot with a fine structure splitting we found QD C, see Fig.~\ref{fig:fig2}c. The difference between the PL components in an applied magnetic field is now simply given by $E(\sigma^+)-E(\sigma^-)$=$\sqrt{\delta_1^2+\Delta_Z^2}$ and we extract a fine structure splitting $\delta_1\approx0.8$~meV and a g-factor of $g\approx-3$, see Fig.~\ref{fig:fig2}f. Here the high value of $\delta_1$ is another signature of the strong Coulomb interaction in these materials \cite{Qiu:2013a,Chernikov:2014a,Echeverry:2016a}. \\
\indent \textit{Charge tuning.---} In Fig.~\ref{fig:fig2}g we show results from the charge tunable sample 3, where the aim is to control the resident carrier density through application of an external bias. Following optical excitation of the suspended part of the flake, applying a bias voltage allows controlling the trion to neutral exciton ratio of the free 2D excitons and also visibly modifies the QD-like emission. Here we focus on a discrete line emitting below the 2D-XT energy in Fig.~\ref{fig:fig2}h.  Starting the experiment at -6V and going towards positive voltage, at an applied voltage of -5 V there appears a clear new feature at lower energy associated with the change of the charge state of the discrete emitter. The energy difference between the charge states is about 2~meV, the discrete jump in emission energy resembles the charge tuning plots observed with self assembled III-V QDs \cite{Warburton:2000a}. Charge control is essential for quantum optics experiments and also optical and electrical manipulation of single spins.  \\
\indent \textit{Perspectives.---}
The discrete emitters in 2D  hosts ML MoSe$_2$ reported here and in WSe$_2$ \cite{Koperski:2015a,Srivastava:2015a,He:2015a,Chakraborty:2015a,Tonndorf:2015a,Kumar:2015a} are a promising platform for quantum optics, as they can be addressed with optical frequencies and placed in tunable micro-cavities \cite{Liu:2015a}. An open question is whether the QD like emitters inherit the valley selection rules. Here the predicted optical selection rules that depend on the size and symmetry of confinement potential will allow in future experiments to aim at identifying the physical origin of the discrete emitters \cite{Liu:2014b,Wu:2016a}.  Concerning spin manipulation, the reduced hyperfine interaction with the nuclear spin bath for states at the K-points (direct gap) as compared to electronic states in III-V semiconductor quantum dots will be an advantage \cite{hyperK,Kormanyos:2014a,Wu:2016a}. The control of the emission properties of these quantum dot like emitters paves the way for further engineering of the light matter interaction in these atomically thin materials.\\
\indent \textit{Acknowledgements.---} We acknowledge funding from ERC Grants No. 306719 and No. 307392, ANR MoS2ValleyControl, and EPSRC grants  EP/I023186/1, EP/K015338/1, and EP/G03673X/1. X.M. acknowledges funding from Institut Universitaire de France and B.D.G. is supported by a Royal Society University Research Fellowship.


\end{document}